\documentclass[doublecol]{epl2}
\usepackage{graphicx}
\usepackage{amsfonts,amsmath,amssymb}
\usepackage[normalem]{ulem}
\usepackage{bbm}

\newcommand{\field}[1]{\mathbb{#1}}

\title{Topological quantum phase transitions of attractive spinless fermions in a honeycomb lattice} 

\author{D. Poletti\inst{1}\!\footnote{Currently at Centre de Physique Th\'eorique, \'Ecole Polytechnique, CNRS, 91128 Palaiseau Cedex, France} \and C. Miniatura\inst{2,1,3} \and B. Gr\'emaud\inst{4,1,3}}
\shortauthor{Dario Poletti \etal}
\institute{
\inst{1} Centre for Quantum Technologies, National University of Singapore, 3 Science Drive 2, Singapore 117543, Singapore\\
\inst{2} Institut Non Lin\'eaire de Nice, UMR 6618, UNS, CNRS; 1361, route des Lucioles, 06560 Valbonne, France\\
\inst{3} Department of Physics, National University of Singapore, 2 Science Drive 3, Singapore 117542, Singapore\\
\inst{4} Laboratoire Kastler-Brossel, Ecole Normale Sup\'erieure, CNRS, UPMC; 4, place Jussieu, 75005 Paris, France} 

\abstract{
We investigate a spinless Fermi gas trapped in a honeycomb optical lattice with attractive nearest-neighbor interactions. At zero temperature, mean-field theory predicts three quantum phase transitions, two being topological. At low interactions, the system is semi-metallic. Increasing the interaction further, the semi-metal destabilizes into a fully gapped superfluid. At larger interactions, a topological transition occurs and this superfluid phase becomes gapless, with Dirac-like dispersion relations. Finally, increasing again the interaction, a second topological transition occurs and the gapless superfluid is replaced by a different fully gapped superfluid phase. We analyze these different quantum phases as the temperature and the lattice filling are varied. 
}


\begin{document}

\maketitle

The discovery of high-$T_c$ superconducting materials\cite{htc,Bonn} has triggered numerous studies to understand the possible underlying pairing mechanisms at work. The seminal work of Bardeen, Cooper, and Schrieffer~\cite{bcs} has thus been extended to more exotic situations, collectively known as unconventional superconductivity, where different types of pairing functions, resonating valence bond states or topological superconductors, to cite a few, have been put under scrutiny~\cite{Anderson,micnas1,micnas2,leermp,Sigrist,Tsuei}. However, if progresses are real, the nagging question of the mechanism of high-$T_c$ superconductivity still remains unanswered to date. In this respect, it might be interesting to look into other related physical systems to get further insights or clues. Indeed, over the past ten years, the advances in the field of ultracold atoms loaded into optical lattices have opened the unprecedented opportunity to study the emergence 
of these possible exotic phases in a highly controllable and accurate way~\cite{Dalibard,Lewenstein_adv_phys,Ketterle}.  Atomic systems are often free of many spurious defects plaguing condensed-matter samples which destroy quantum coherence. Furthermore, the interaction strength between atoms, relative to their tunneling amplitude, can be tuned over orders of magnitude, be it with the help of Feshbach resonances \cite{Timmermans99} or by increasing the optical lattice depth, while the lattice geometry is perfectly under command. This is exemplified by the Mott-Superfluid transition observed with ultracold atoms \cite{Bloch02, Bloch08, Esslinger08}. This interaction energy can even be turned attractive or repulsive, a key feature in the experimental studies of the celebrated BEC-BCS crossover \cite{Jin03,Jin04,Salomon04}.
Recently, with the laboratory realization in 2004 of graphene sheets \cite{Novoselov04}, the honeycomb lattice has attracted a lot of attention as its low-energy 
excitations around half-filling behave as massless Weyl-Dirac fermions. This situation could be easily mimicked with ultracold atoms~\cite{Blakie,Duan,cqt} where different physical models of attractive fermions in a honeycomb lattice have been analyzed and different surperfluid states 
have been proposed \cite{paramekanti,Doniach, Neto, LeHur, Herbut}. In particular, different ways of producing nearest-neighbor interactions have been studied in \cite{Neto} for graphene and in \cite{LeHur, Wang, Lewenstein, Jaksch, Jiang, Morais, Cooper} for ultracold gases where large values seem actually reachable for composite fermions in Fermi-Bose mixtures \cite{Massignan}. 
For instance, nearest-neighbor interaction strengths as large as $V\sim6t$ (where $t$ is the nearest-neighbor hopping amplitude) have been reported for the $^{171}{\rm Yb}-^{174}\!{\rm Yb}$ mixture at zero magnetic field  \cite{julienne, Massignan}. Another promising candidate is the $^6{\rm Li}-^7\!{\rm Li}$ mixture for which $V$ can be tuned using homonuclear and heteronuclear s-wave Feshbach resonances \cite{sjjmf}. 

In this Letter, we consider a one-component fermionic gas loaded on the honeycomb lattice with nearest-neighbor attractive interactions. Using a mean-field treatment, we show that this system undergoes three first-order phase transitions at zero temperature, two of which are topological, as the interaction strength is increased. In particular the system jumps from a fully gapped superfluid (SF) phase to a gapless one (one with zero-energy excitation modes) and back again to a fully gapped SF phase.

\begin{figure}[!h] 
\center
\includegraphics[scale=0.9]{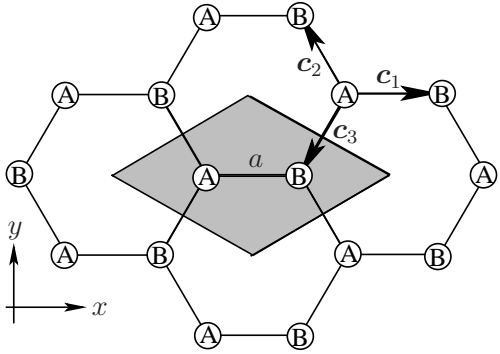} 
\caption{The Bravais lattice of a two-dimensional regular honeycomb lattice is a triangular lattice with a two-point basis cell (grey-shaded diamond-shaped area with points $A$ and $B$). The mean-field order parameter $\boldsymbol \delta \in\field{C}^3$ is defined
by the components $\delta_\alpha = \langle \hat{b}_j  \hat{a}_i \rangle$ ($\alpha=1,2,3$), $i$ and $j$ being nearest-neighbor sites connected by  ${\bf c}_\alpha$. In the paper, we set the lattice constant $a=|{\bf c}_\alpha|$ to unity.} \label{honeylatt}  
\end{figure}

The honeycomb lattice consists of two shifted triangular sublattices, one labeled with $A$ sites and the other with $B$ sites. Each $A$ site is connected to its three adjacent $B$ sites by the vectors ${\bf c}_\alpha$ ($\alpha=1,2,3$) and the honeycomb diamond-shaped Bravais unit cell contains exactly one $A$ site and one $B$ site, see fig.~\ref{honeylatt}. The Fermi-Hubbard tight-binding Hamiltonian of our one-component system reads

\begin{equation} 
\label{Full-H}
\hat{H} = -t\sum_{\langle i,j\rangle} \left(\hat{a}^{\dagger}_i \hat{b}_j+\hat{b}^{\dagger}_j\hat{a}_i\right) - V\sum_{\langle i,j\rangle}  \hat{a}^{\dagger}_i \hat{b}^{\dagger}_j \hat{b}_j \hat{a}_i  ,
\end{equation} 
where $t$ is the hopping amplitude, and $V$ the strength of the nearest-neighbor interaction. The fermionic operators $\hat{a}_i$ and $\hat{b}_j$ annihilate a fermion on $A$ and $B$ sites respectively and $\langle i,j\rangle$ restricts the corresponding sums to nearest-neighbor sites. Throughout the paper we use $t$ as the energy scale  and set $t=1$ hereafter.

\begin{figure}[!h] 
\center
\includegraphics[width=\columnwidth ]{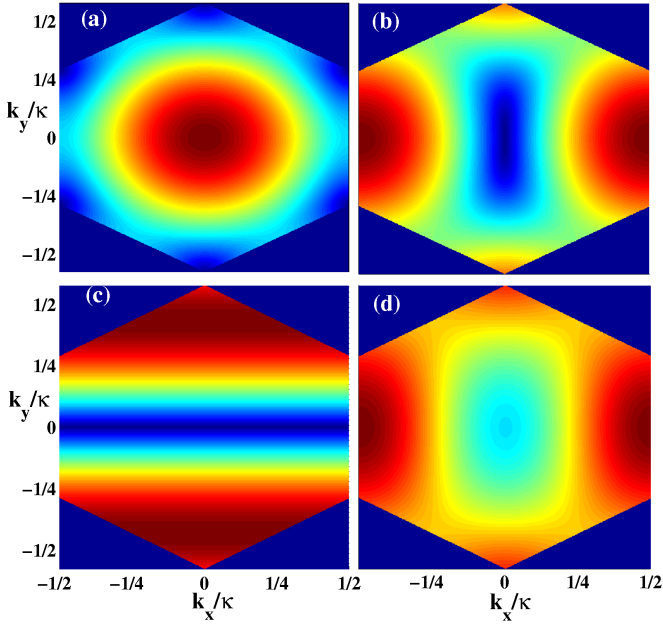} 
\caption{(color online) Plot of $|\Delta_{\bf k}|$ in the hexagonal first Brillouin zone of the honeycomb lattice as obtained for (a) ${\boldsymbol \delta}={\bf u}_{1}$, (b) ${\boldsymbol \delta}={\bf u}_{2}$, (c) ${\boldsymbol \delta}={\bf u}_{3}$. The interaction strength has been artificially set to $V=1$. The Bloch wave vector ${\bf k}$ has been expressed in units of $\kappa = 4\pi/(3a)$ so that the ranges in the hexagonal Brillouin zone are $|k_x/\kappa|\leq 1/2$ and $|k_y/\kappa| \leq 1/\sqrt{3}$ respectively. Plot (d) is obtained from the actual minimization of the free-energy at $V=11$ (${\boldsymbol \delta}$ is a  certain linear combination of ${\bf u}_{1}$ and ${\bf u}_{2}$). In this last figure the color scale has been modified to help visualize the pattern.} \label{AbsDelt}  
\end{figure}

Dropping the Hartree-Fock terms, the mean-field Hamiltonian in Fourier space reads
\begin{equation} 
\begin{array}c 
{\displaystyle \hat{\mathcal{H}}_{M\!F}=-\sum_{{\bf k}}\left( \gamma^*_{{\bf k}} \hat{a}^{\dagger}_{{\bf k}}  \hat{b}_{{\bf k}} +\gamma_{{\bf k}}  \hat{b}^{\dagger}_{{\bf k}}  \hat{a}_{{\bf k}}   \right) 
+ } \\ 
{\displaystyle  \sum_{{\bf k}}  \left(\Delta^*_{{\bf k}} \; \hat{b}_{-{\bf k}}  \hat{a}_{{\bf k}} + \Delta_{{\bf k}}\; \hat{a}^{\dagger}_{{\bf k}}  \hat{b}^{\dagger}_{-{\bf k}} -\Delta_{{\bf k}}\; \langle \hat{a}^{\dagger}_{{\bf k}}  \hat{b}^{\dagger}_{-{\bf k}} \rangle \right) }  ,
\end{array} \label{H_MF}
\end{equation} 
where $\gamma_{{\bf k}}=\sum_l e^{{\bf i} {\bf k}\cdot {\bf c}_l  }$, $\langle \cdot \rangle$ denotes the grand-canonical statistical average and ${\bf k}$ spans the honeycomb first Brillouin zone, a regular hexagon with area $\Omega=8\pi^2/(3\sqrt{3}a^2)$.
The pairing field $\Delta_{{\bf k}}$ reads
\begin{equation}
{\displaystyle \Delta_{{\bf k}}= -\frac {V} {N_c}  \sum_{{\bf k}'}   \gamma_{{\bf k}-{\bf k}'} \  \langle \hat{b}_{-{\bf k}'}  \hat{a}_{{\bf k}'}\rangle },
\end{equation}
with $N_c$ the number of lattice unit cells.
We define the order parameter $\boldsymbol \delta \in\field{C}^3$
with components $\delta_\alpha = \langle \hat{b}_j  \hat{a}_i \rangle$ ($\alpha=1,2,3$), $i$ and $j$ being nearest-neighbor sites connected by  ${\bf c}_\alpha$, see fig.~\ref{honeylatt} \cite{Jiang,Morais}. Due to the $U(1)$ invariance of $\hat{H}$, $\boldsymbol \delta$ is defined up to a global phase. As $\hat{H}$ is left invariant by the point group $C_{3v}$, any permutation of the components of ${\boldsymbol \delta}$ provides another mean-field solution: the right and left $2\pi/3$ rotations correspond to right and left cyclic permutations of the $\delta_\alpha$ and the reflections about $\textbf{c}_\alpha$ to transpositions. The permutation group $\mathcal{S}_3$ splits the order parameter space $\field{C}^3$ into a direct sum of two orthogonal invariant subspaces. One is spanned by 
${\bf u}_{1}\!=\!\frac 1 {\sqrt{3}}(1, 1, 1)$ and corresponds to the symmetric one-dimensional irreducible  representation of $\mathcal{S}_3$. The other one is spanned by ${\bf u}_{2}\!=\!\frac 1 {\sqrt{6}}(2, -1, -1) $ and 
${\bf u}_{3}\!=\!\frac 1 {\sqrt{2}}(0, 1, -1)$ (this particular choice of basis vectors will become clear later) and corresponds to the two-dimensional irreducible representation of $\mathcal{S}_3$. Writing ${\boldsymbol \delta}=\sum_\alpha \eta_\alpha {\bf u}_{\alpha}$, we use the gap parameter norm $\delta = \sqrt{\sum_\alpha |\eta_\alpha|^2}$ and the relative weights $w_\alpha= |\eta_\alpha |/\delta$ to quantify the strength and geometry of the SF order. We find
\begin{eqnarray} 
\Delta_{{\bf k}} &=&  -V {\boldsymbol \delta} \cdot {\bf f}({\bf k}) = -V \sum_\alpha \eta_\alpha f_\alpha ,
\end{eqnarray} 
with 
\begin{eqnarray}
 f_1 &=& \frac 1 {\sqrt{3}}\gamma_{-{\bf k}} \\ \nonumber
f_2 &=&\frac 1 {\sqrt{6}}(2e^{- {\bf i}  {\bf k}\cdot {\bf c}_1 }-e^{- {\bf i}  {\bf k}\cdot {\bf c}_2 }-e^{- {\bf i}  {\bf k}\cdot {\bf c}_3 })\\
f_3 &=&\frac 1 {\sqrt{2}}(e^{- {\bf i}  {\bf k}\cdot {\bf c}_2 }-e^{- {\bf i}  {\bf k}\cdot {\bf c}_3 }). \nonumber
\end{eqnarray}
We note that ${\bf f}^*({\bf -k}) = {\bf f}({\bf k})$. The modulus of the gap parameter $|\Delta_{\mathbf{k}}|$ in the Brillouin zone is depicted in figs.\ref{AbsDelt}(a-c) for each of the geometries ${\bf u}_\alpha$ at $T=0$ and $\mu=0$. We next diagonalize the grand-canonical mean-field Hamiltonian $\hat{\mathcal{H}}= \hat{\mathcal{H}}_{MF}-\mu \hat{N}$ through a Bogoliubov-Valatin transformation and obtain
\begin{eqnarray} 
\hat{\mathcal{H}}&=& N_c \, F_0 +  \sum_{{\bf k},s=\pm} E_{s}({\bf k})  \hat{c}^{\dagger}_{{\bf k}s}\hat{c}_{{\bf k}s}\\
F_0 &=& -\mu + V \delta^2  - \frac 1 {2N_c} \sum_{{\bf k},s=\pm}  E_s({\bf k}).
\end{eqnarray} 
The operators $\hat{c}_{{\bf k}s}$ annihilates a Bogoliubov quasi-particle with momentum ${\bf k}$ and index $s$ and $\mu$ is the chemical potential ($\mu = 0$ at half-filling).
The excitation spectrum is given by ${\displaystyle E_{\pm} ({\bf k}) =  \sqrt{ X  \pm \sqrt{ Y}   } } $ where
\begin{eqnarray}
 X &=& \mu^2 +|\gamma_{{\bf k}}|^2 + \left(|\Delta_{{\bf k}}|^2 +|\Delta^*_{-{\bf k}}|^2\right)/2\\ \nonumber
Y &=& \left(|\Delta_{{\bf k}}|^2 - |\Delta^*_{-{\bf k}}|^2\right)^2/4 + 
2\Re e\!\left( \Delta_{-{\bf k}}^* \Delta_{{\bf k}}\gamma_{{\bf k}}^{2}\right)\\ \nonumber
 && +  |\gamma_{{\bf k}} |^2 (4\mu^2 + |\Delta_{{\bf k}}|^2  + |\Delta^*_{-{\bf k}}|^2 ).
\end{eqnarray}
 The inversion symmetry exchanging the $A$ and $B$ sublattices in $\hat{H}$ leads to $E_s({\bf k})=E_s({\bf -k})$ in $\hat{\mathcal{H}}$ and to the innocuous change $\boldsymbol \delta \to -\boldsymbol \delta$, already covered by the $U(1)$ invariance.
The spectrum is also invariant under $\Delta_{{\bf k}}  \rightarrow \Delta^*_{{\bf -k}}$ or, equivalently, under ${\boldsymbol \delta}\rightarrow {\boldsymbol \delta^*}$. 
This symmetry reflects the time reversal invariance of $\hat{H}$ and implies that both ${\boldsymbol \delta}$ and its time-reversed partner ${\boldsymbol \delta}^*$ are mean-field solutions of $\hat{\mathcal{H}}$. If ${\boldsymbol \delta}$ and ${\boldsymbol \delta}^*$ cannot be matched by the $U(1)$ symmetry, i.e. if ${\boldsymbol \delta}$ is genuinely complex, then the system exhibits spontaneous time-reversal symmetry breaking superfluidity.

The free energy per unit cell at temperature $T$ is given  by 
\begin{eqnarray} 
F = - \frac {\ln Z} {\beta N_c}   =   F_0-\frac 1 {\beta N_c} \sum_{{\bf k},s=\pm}  \ln\left(1+e^{-\beta E_s({\bf k})}\right) ,
\label{eqF}  
\end{eqnarray} 
where $\beta=1/(k_B T)$ is the inverse temperature, $k_B$ being the Boltzmann constant. $F_0$ identifies with the free energy per unit cell at $T=0$. To compute the pairing field, it proves numerically convenient to directly find the minima of $F$ as a function of ${\boldsymbol \delta}$ for given values of  $V$, $\mu$ and $\beta$ and extract the corresponding values of the order parameter components $\eta_\alpha$. In our computations, the $\eta_1$ component was kept real but the two other complex. An alternative equivalent procedure is to solve the three coupled gap equations obtained from $(\partial F/\partial \eta_\alpha^*)({\boldsymbol \delta},\mu,V,\beta) = 0$ ($\alpha=1,2,3$). In the thermodynamic limit, they read
\begin{equation}
\eta_\alpha = \sum_{s=\pm} \int \!d{\bf k} \ \frac{\tanh(\beta E_s/2)}{2V\Omega} \ \frac{\partial E_s}{\partial \eta^*_\alpha} .
\label{self-const}    
\end{equation} 
In all our calculations, the free-energy minimum was always found for a real ${\boldsymbol \delta}$, ruling out time-reversal symmetry breaking in our system. 

\begin{figure}[!h] 
\center
\includegraphics[width=\columnwidth ]{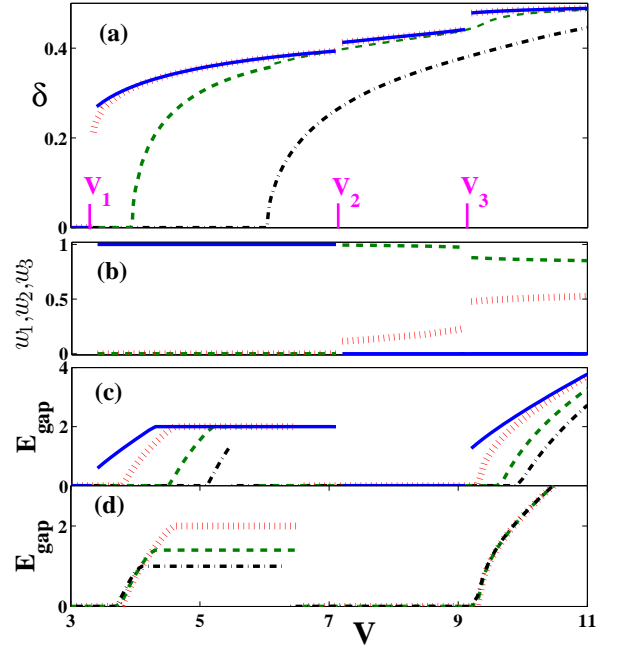} 
\caption{(color online) (a) Plot of the order parameter strength $\delta$ vs $V$ for $\mu=T=0$ (blue continuous line), $\beta=5$ (red dotted line), $\beta=1.5$ (green dashed line) and $\beta=0.75$ (black dash-dotted line). (b) Plot of the geometry weights $w_1$ (red dotted line), $w_2$ (green dashed line) and $w_3$ (blue continuous line) at $\mu=T=0$. (c) Energy gap $E_{{\rm gap}}$ at $\mu=0$ versus interaction strength $V$ for $T=0$ (blue continous line), $\beta=2$ (red dotted line), $\beta=1.25$ (green dashed line) and $\beta=1$ (black dot-dashed line). (d) Excitation gap $E_{{\rm gap}}$ at $\beta=2$ versus interaction strength $V$ for $\mu=0$ (red dotted line), $\mu=0.3$ (green dashed line) and $\mu=0.5$ (black dot-dashed line).}  \label{gapanalysis} 
\end{figure}

We plot in fig.\ref{gapanalysis}(a) the order parameter norm $\delta$ at $\mu=0$ as a function of $V$. At $T=0$ (blue solid line), we observe three discontinuous jumps signaling first-order quantum phase transitions.The first one occurs at $V_1\approx 3.36$, the second one at $V_2\approx 7.12$ and the third one at $V_3\approx 9.15$. We observed that, for $V_1<V<V_2$ and up to permutations, one of the components $\delta_\alpha$ is always zero while the other two are always opposite of each other. For $V>V_2$, and again up to permutations, two of the components $\delta_\alpha$ are always identical, the remaining one being different. This means that, for $V_1<V<V_2$, the order parameter can always be recast in the form ${\boldsymbol \delta}=\eta_3 {\bf u}_3$, while for  $V>V_2$ it always reads ${\boldsymbol \delta}=\eta_1 {\bf u}_1+\eta_2 {\bf u}_2$. The three relative weights $w_\alpha$ at $T=0$ and $\mu=0$ are plotted in fig.\ref{gapanalysis}(b). We further note that the dominant geometry is  ${\bf u}_{2}$ in the range $V_2 < V < V_3$ since $w_2 \approx 1$, meaning that $\delta_2=\delta_3 \approx -\delta_1/2$. However, at $V=V_3$, the weight of ${\bf u}_{1}$ abruptly increases from $w_1 \approx 0.25$ to $w_1 \approx 0.5$. For $V>V_3$, one of the $\delta_\alpha$ largely dominates over the other two and ${\boldsymbol \delta}$ is essentially along ${\bf c}_\alpha$. In fig.\ref{AbsDelt}(d) we have plotted $|\Delta_{\mathbf{k}}|$ for the particular case $V=11$.

\begin{figure}[!h] 
\center
\includegraphics[width=\columnwidth ]{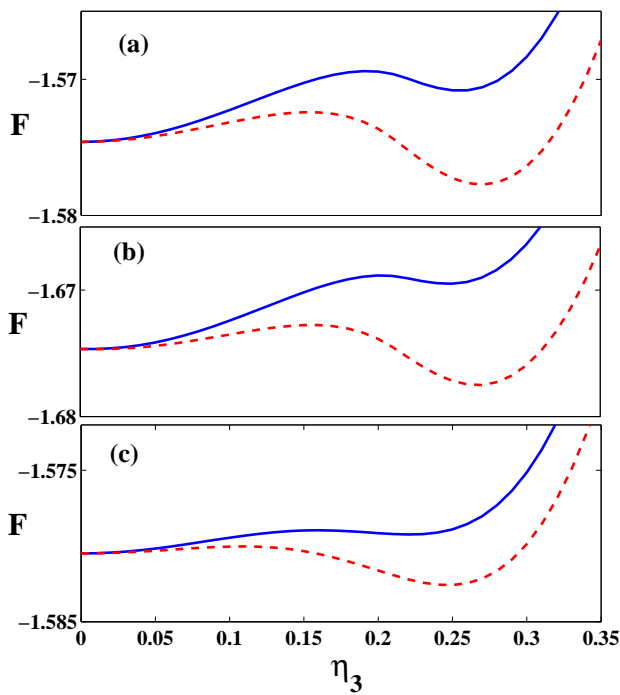} 
\caption{(color online) Plot of the free energies $F$ versus $\eta_3$ ($\eta_1=\eta_2=0$) below the transition at $V_1$ (semi-metallic regime) and above $V_1$ (superfluid regime). (a) $T=0$, $\mu=0$, $V=3.3$ (continuous line) and $V=3.4$ (dashed line); (b) $T=0$, $\mu=0.1$, $V=3.26$ (continuous line) and $V=3.38$ (dashed line); (c) $\beta=5$, $\mu=0$, $V=3.35$ (continuous line) and $V=3.42$ (dashed line). The local minimum at $\eta\neq 0$ (continuous line) becomes a global one (dashed line) leading to a jump of the order parameter. Therefore the transition at $V_1$ is of the first-order also for a moderate increase of the temperature as well as a departure from half-filling.} \label{FirstOrder}   
\end{figure}

The first order nature of the transition between the semi-metallic and the superfluid phase, revealed by a jump in the order parameter (also for $T>0$ and $\mu\neq 0$) is further corroborated by fig.\ref{FirstOrder}. Here we have plotted the free energy $F$ for interaction strength corresponding to either the semi-metallic regime (continuous curves) or the superfluid regime (dashed curves) against $\eta_3$ (which is the relevant parameter to describe the transition as clear from the previous discussion) for different values of $\beta$ and $\mu$. More precisely the three plots correspond to (a) $T=0$ and $\mu=0$ (b) $T=0$ and $\mu=0.1$ and lastly (c) $\beta=5$ and $\mu=0$. 
For these typical cases, the free energy clearly presents two well separated minima for each curve: one at $\eta_3=0$ and one at $\eta\neq 0$. The global minima determines the groundstate of the system while the local minima corresponds to a metastable state. When crossing the critical interaction strength (i.e. when the two minima are the same), the order parameter ($\eta_3$) of the groundstate will jump abruptly between $0$ and a finite value. Hence, for these different cases, including those with $T>0$ and $\mu\neq 0$, the system undergoes first order phase transitions.     

A deeper insight into the various emerging solutions can be obtained from group theoretic arguments  by expanding the gap equations \eqref{self-const} near ${\boldsymbol\delta}={\bf 0}$ up to second order in $\delta$~\cite{Sigrist}. The linear term in this Ginzburg-Landau analysis of the gap equations is characterized by a diagonal $3\times3$ matrix $M_0$ that is split into the previous irreducible representations of $\mathcal{S}_3$. In our case, the SF onset is obtained when the doubly degenerate eigenvalue of $M_0$ is unity, thus selecting two dominant SF orders. At this linear level, the SF order can a priori develop into any superposition of ${\bf u}_{2}$ and ${\bf u}_{3}$.
However the subsequent nonlinear terms lift this degeneracy and only specific combinations of the two SF orders actually survive~\cite{Sigrist}. Assuming the SF order starts in only one of these dominant geometries, the question arises whether a sub-dominant SF order could simultaneously develop, as $T$ or $V$ change, without the need for an additional phase transition. For this to happen, the sub-dominant order must belong to the same irreducible representation of  the (lower) symmetry group leaving invariant the excitation energies associated to the dominant SF order~\cite{Sigrist}.
In our case,  ${\bf u}_2$ is invariant under swapping its last two components, and so is ${\bf u}_1$, whereas ${\bf u}_3$ flips sign. Hence a sub-dominant ${\bf u}_1$-SF order can only develop along with ${\bf u}_2$. In this scenario, the Ginzburg-Landau theory predicts that the coefficient  $\eta_2$ of the dominant geometry ${\bf u}_{2}$ vanishes like $(T_c-T)^{1/2}$ near the critical temperature, whereas the coefficient  $\eta_1$ of the sub-dominant one ${\bf u}_{1}$ vanishes like $(T_c-T)^{3/2}$~\cite{Sigrist}. We have numerically verified this prediction. In addition the transition between the ${\bf u}_3$-SF order and the mixture of the ${\bf u}_2$ and ${\bf u}_1$ SF orders must result from two successive second-order phase transitions or from a first-order one, the latter being the one observed at $V=V_2$.

Fig.\ref{gapanalysis}(a) also shows how $\delta$ varies with $V$ as $T$ is varied. At $\beta=5$ (red dotted line), the physics is qualitatively the same as at $T=0$. At sufficiently high $T$, e.g. $\beta=1$ (green dashed line), the first and third transitions smoothen and become of higher order. The second transition, which implies a radical change in the geometry of ${\boldsymbol \delta}$, is always first-order as long as ${\bf u}_{3}$ persists. Above some $T$ ($\beta=0.75$), ${\bf u}_{3}$ no longer appears.  

The nature of each SF order is inferred from the energy gap $E_{{\rm gap}}= \min_{{\bf k}}\left(2E_{-}({\bf k}) \right)$, i.e. from the minimum energy needed to create an excitation in the system. From fig.\ref{gapanalysis}(c), we see that at $\mu=0$ and $T=0$ the SF order is fully gapped below $V_2$, gapless for $V_2<V<V_3$ and again fully gapped above $V_3$; the transitions at $V_2$ and $V_3$ are thus TQPTs \cite{Melo1,Melo2}. The Chern number \cite{hatsugai,Chern1,Chern2} of the bands is zero and, given the symmetry of the Hamiltonian, the zero-energy modes appear in pairs and are not topologically protected \cite{Gurarie}. We next verified that these TQPTs persist at higher $T$ and away from half-filling. Fig.\ref{gapanalysis}(c) shows $E_{{\rm gap}}$ at $\mu=0$ for different $T$ as $V$ increases. 
For $T$ as large as at $\beta=1$, we still observe three distinct SF regions, their specific ranges in $V$ depending on $T$. Fixing $T$ at the experimentally attainable value $\beta=2$, fig.\ref{gapanalysis}(d) gives $E_{{\rm gap}}$ for different chemical potentials. As large values of $V$ could be achieved \cite{Massignan}, the observation of these two TQPTs with ultracold gases seems within experimental reach in the near future.

\begin{figure}[!h] 
\center
\includegraphics[width=\columnwidth ]{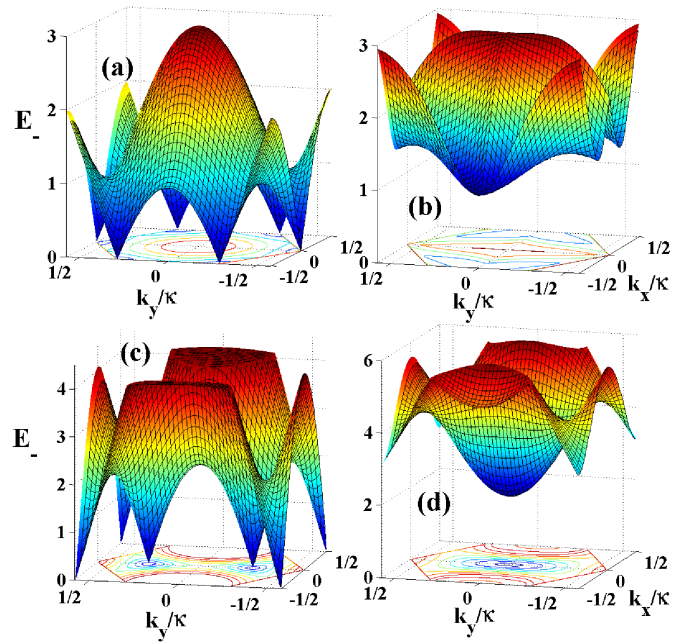} 
\caption{ (color online) Plots of $E_{-}({\bf k})$ in the Brillouin zone at $\mu=T=0$. (a) Honeycomb lattice tight-binding band structure $|\gamma_{{\bf k}}|$ at $V=0$. (b) Fully-gapped SF phase at $V=5$. (c) Gapless SF phase at $V= 8$. (d) Fully gapped SF phase at $V=11$. The Bloch wave vector ${\bf k}$ has been expressed in units of $\kappa=4\pi/(3a)$ so that the ranges in the hexagonal Brillouin zone are $|k_x/\kappa|\leq 1/2$ and $|k_y/\kappa| \leq 1/\sqrt{3}$ respectively.}  \label{E2} 
\end{figure}

Fig.\ref{E2} gives further evidence of the dramatic changes occurring in the excitation spectrum $E_{-}({\bf k})$ in the Brillouin zone when the different SF orders are achieved by increasing $V$ at $T=0$ and $\mu=0$. In the semi-metallic region, see fig.\ref{E2}(a), one recovers the usual honeycomb lattice tight-binding band structure when $V\to0$. Fig.\ref{E2}(b) gives the spectrum for the fully gapped SF phase at $V=5$. At $V=8$, a gapless SF order develops and $E_{-}({\bf k})$ presents two zeros with linear dispersion, see fig.\ref{E2}(c). The tips of these cones move closer to each other when $V$ is increased but do not merge at $T=0$. Instead, the spectrum changes abruptly at $V_3$ and displays a single nonzero minimum, signaling again a fully gapped SF order. This is shown in fig.\ref{E2} for $V=11$. At higher $T$, this transition becomes smooth. The two cones merge into one single zero energy minimum with linear dispersion in one direction and quadratic in the orthogonal one. A similar-looking topological transition is obtained for non-interacting fermions in graphene when the nearest-neighbor hopping amplitudes are imbalanced \cite{Goerbig1,Goerbig2}. The energy gap $E_{{\rm gap}}$ then increases smoothly from zero when $V$ is further increased and the system enters again a fully gapped SF order.

As a conclusion, we have observed and studied two topological quantum phase transitions in a one-component Fermi gas loaded in a honeycomb lattice with nearest-neighbour attractive interactions.
The corresponding SF features are robust against moderate changes in the temperature and in the chemical potential. These topological transitions should be within the reach of ultracold gases experiments. The different pairing geometries achieved by the system could be analyzed by observing the momentum distribution of the expanding gas after release from the trap. 

To estimate the critical temperature of the transition (Kosterliz-Thouless-like) between the gas of paired fermions and the superfluid phase, one must take into account the gaussian fluctuations beyond the saddle-point mean-field approximation~\cite{Melo1,Melo2,paramekanti}. This will be undertaken in a future work.   

Centre for Quantum Technologies is a Research Centre of Excellence funded by the Ministry of Education and National Research Foundation of Singapore. This work has been supported by the CNRS PICS 4159 (France) and by the France-Singapore Merlion program (FermiCold 2.01.09).
The authors thank C. A. R. S\'a de Melo, S. Das Sarma, P. Massignan and B.-G. Englert for useful discussions.


\begin{thebibliography}{99} 
\bibitem{htc} {\sc Bednorz} J.G. and {\sc M\"uller} K.A., {\it Z. Phys. B}, \textbf{64} (1986) 189.
\bibitem{Bonn} {\sc Hardy} W.N., {\sc Bonn} D.A., {\sc Morgan} D.C., {\sc Liang} R. and {\sc Zhang} K., {\it Phys. Rev. Lett.}, {\bf 70} (1993) 3999.  
\bibitem{bcs} {\sc Bardeen} J., {\sc Cooper} L.N., and {\sc Schrieffer} J.R., {\it Phys. Rev.}, \textbf{108} (1957) 1175.
\bibitem{Anderson} {\sc Anderson} P.W. and {\sc Morel} P., {\it Phys. Rev}, {\bf 123} (1961) 1911. 
\bibitem{micnas1} {\sc Micnas} R., {\sc Ranninger} J., {\sc Robaszkiewicz} S., and {\sc Tabor} S., {\it Phys. Rev. B}, \textbf{37} (1988) 9410.
\bibitem{micnas2} {\sc Micnas} R., {\sc Ranninger} J., and {\sc Robaszkiewicz} S., {\it Rev. Mod. Phys.}, \textbf{62} (1990) 113.
\bibitem{leermp} {\sc Lee} P.A., {\sc Nagaosa} N., and {\sc Wen} X.-G., {\it Rev. Mod. Phys.}, \textbf{78} (2006) 17.
\bibitem{Sigrist} {\sc Sigrist} M. and {\sc Ueda} K., {\it Rev. Mod. Phys.}, {\bf 63} (1991) 239.
\bibitem{Tsuei} {\sc Tsuei} C.C. and {\sc Kirtley} J.R., {\it Rev. Mod. Phys.}, \textbf{72} (2000) 969.
\bibitem{Dalibard} {\sc Bloch} I., {\sc Dalibard} J. and {\sc Zwerger} W., {\it Rev. Mod. Phys}, {\bf 80} (2008) 885.
\bibitem{Lewenstein_adv_phys} {\sc Lewenstein} M., {\sc Sanpera} A., {\sc Ahufinger} V., {\sc Damski} B., {\sc Sen} Aditi, and {\sc Sen} U., {\it Adv.~Phys.}, \textbf{56} (2007) 243.
\bibitem{Ketterle} {\sc Ketterle} W. and {\sc Zwierlein} M., {\it Proceedings of the International School of Physics ``Enrico Fermi''},  \textbf{164}, edited by {\sc Inguscio} M., {\sc Ketterle} W. and {\sc Salomon C.} (IOS Press, Amsterdam) 2007, pp. 95-287.
\bibitem{Timmermans99} {\sc Timmermans} E., {\sc Tommasini} P., {\sc Hussein} M., and {\sc Kerman} A., {\it Phys. Rep.}, {\bf 315} (1999) 199.
\bibitem{Bloch02} {\sc Greiner} M., {\sc Mandel} O., {\sc Esslinger} T., {\sc H\"ansch} T. W., and {\sc Bloch} I., {\it Nature}, {\bf 415} (2002) 39.
\bibitem{Bloch08} {\sc Schneider} U. {\it et al.}, {\it Science}, {\bf 322} (2008) 1520.
\bibitem{Esslinger08} {\sc J\"ordens} R., {\sc Strohmaier}, {\sc G\"unter} K., {\sc Moritz} H. and {\sc Esslinger} T., {\it Nature}, {\bf 455} (2008) 204.
\bibitem{Jin03} {\sc Greiner} M., {\sc Regal} C. A., and {\sc Jin} D. S., {\it Nature}, {\bf 426} (2003) 537.
\bibitem{Jin04} {\sc Regal} C. A., {\sc Greiner} M., and {\sc Jin} D. S., {\it Phys. Rev. Lett.}, {\bf 92} (2004) 040403.
\bibitem{Salomon04} {\sc Bourdel} T. {\it et al.}, {\it Phys. Rev. Lett.}, {\bf 93} (2004) 050401. 
\bibitem{Novoselov04} {\sc Novoselov} K. S. {\it et al.}, {\it Science}, {\bf 306} (2004) 666.
\bibitem{Blakie} {\sc Blakie P.B.} and {\sc Clark C.W.}, {\it J. Phys. B} {\bf 37}  (2004) 1391. 
\bibitem{Duan} {\sc Zhu} S.-L., {\sc Wang} B. and {\sc Duan} L.-M., {\it Phys. Rev. Lett.}, {\bf 98} (2007) 260402. 
\bibitem{cqt} {\sc Lee} K.L., {\sc Gr\'emaud} B., {\sc Han} R., {\sc Englert} B.-G. and {\sc Miniatura} C., {\it Phys. Rev. A}, {\bf 80} (2009) 043411. 
\bibitem{paramekanti} {\sc Zhao} E. and {\sc Paramekanti} A., {\it Phys. Rev. Lett.}, {\bf 97} (2006) 230404.  
\bibitem{Doniach} {\sc Black-Schaffer} A.M. and {\sc Doniach} S., {\it Phys. Rev. B}, {\bf 75} (2007) 134512. 
\bibitem{Neto} {\sc Uchoa} B. and {\sc Castro Neto} A.H., {\it Phys. Rev. Lett.}, {\bf 98} (2007) 146801. 
\bibitem{LeHur} {\sc Bergman} D.L. and {\sc Le Hur} K., {\it Phys. Rev B}, {\bf 79} (2009) 184520. 
\bibitem{Herbut} {\sc Roy} B. and {\sc Herbut} I.F., {\it Phys. Rev. B}, {\bf 82} (2010) 035429. 
\bibitem{Wang} {\sc Wang} D.-W., {\sc Lukin} M.D. and {\sc Demler} E., {\it Phys. Rev. A}, {\bf 72} (2005) 051604(R).
\bibitem{Lewenstein} {\sc Lewenstein} M., {\sc Santos} L., {\sc Baranov} M.A. and {\sc Fehrmann} H., {\it Phys. Rev. Lett.}, {\bf 92} (2004) 050401.
\bibitem{Jaksch} {\sc Bruderer} M., {\sc Klein} A., {\sc Clark} S. R., and {\sc Jaksch} D., {\it Phys. Rev. A}, {\bf 76} (2007) 011605(R).
\bibitem{Jiang} {\sc Jiang} Y., {\sc Yao}, {\sc Carlson} E. W., {\sc Chen} H.-D. and {\sc Hu} J.-P., {\it Phys. Rev. B}, {\bf 77} (2008) 235420.
\bibitem{Morais} {\sc Lim} L.K., {\sc Lazarides} A., {\sc Hemmerich} A. and {\sc Morais-Smith} C., {\it EuroPhys. Lett.}, {\bf 88} (2009) 36001.
\bibitem{Cooper} {\sc Cooper} N. R. and {\sc Shlyapnikov} G., {\it Phys. Rev. Lett.}, {\bf 103} (2009) 155302.
\bibitem{Massignan} {\sc Massignan} P., {\sc Sanpera} A. and {\sc Lewenstein} M., {\it Phys. Rev. A}, {\bf 81} (2010) 031607(R).
\bibitem{julienne} {\sc Kitagawa M.} {\it et al.}, {\it Phys. Rev. A}, {\bf77} (2008) 012719.
\bibitem{sjjmf} {\sc v. Kempen E.G.M., Marcelis B.} and {\sc Kokkelmans S.J.J.M.F.}, {\it Phys. Rev. A}, {\bf 70} (2004) 050701(R). 
\bibitem{Melo1} {\sc Botelho} S.S. and {\sc Sa de Melo} C.A.R., {\it Phys. Rev. B}, {\bf 71} (2005) 134507.
\bibitem{Melo2} {\sc Iskin} M. and {\sc Sa de Melo} C.A.R., {\it Phys. Rev. A}, {\bf 76} (2007) 013601.
\bibitem{hatsugai} {\sc Hatsugai} Y. and {\sc Ryu} S., {\it Phys. Rev. B}, \textbf{65} (2002) 212510. 
\bibitem{Chern1} {\sc Nagaosa} N., {\sc Sinova} J., {\sc Onoda} S., {\sc MacDonald} A. H. and {\sc Ong} N. P., {\it Rev. Mod. Phys}, {\bf 82} (2010) 1539.
\bibitem{Chern2} {\sc Cheng} M., {\sc Sun} K., {\sc Galitski} V. and {\sc Das Sarma} S., {\it Phys. Rev. B}, {\bf 81} (2010) 024504.
\bibitem{Gurarie} {\sc Gurarie} V. and {\sc Radzihovsky} L., {\it Phys. Rev. B}, {\bf 75} (2007) 212509. 
\bibitem{Goerbig1} {\sc Montambaux} G., {\sc Pi\'echon} F., {\sc Fuchs} J.-N., and {\sc Goerbig} M.O., {\it Europhys. J. B}, {\bf 72} (2009) 509. 
\bibitem{Goerbig2} {\sc Montambaux} G., {\sc Pi\'echon} F., {\sc Fuchs} J.-N. and {\sc Goerbig} M.O., {\it Phys. Rev. B}, {\bf 80} (2009) 153412.

\end{thebibliography}
\end{document}